\documentclass[pra,aps,twocolumn,longbibliography,nobalancelastpage, superscriptaddress]{revtex4-1}
\usepackage[english]{babel}
\usepackage[utf8]{inputenc}
\usepackage{xcolor}
\usepackage{svg}
\usepackage{dsfont}
\usepackage{subfig}
\usepackage{float}
\usepackage{graphicx}
\usepackage{tablefootnote}
\usepackage{footnote}
\usepackage{soul}
\usepackage{dsfont}
\usepackage{mathrsfs,amsmath}
\usepackage[separate-uncertainty=true,multi-part-units=single]{siunitx}
\usepackage{amsthm}
\usepackage{mathtools}
\usepackage{physics}
\usepackage{xcolor}
\usepackage{graphicx}
\usepackage{multirow}
\usepackage[left=23mm,right=13mm,top=35mm,columnsep=15pt]{geometry} 
\usepackage{adjustbox}
\usepackage{placeins}
\usepackage[T1]{fontenc}
\usepackage{lipsum}
\usepackage{csquotes}
\usepackage{ulem}
\usepackage{subfig}

\usepackage[pdftex, pdftitle={Article}, pdfauthor={Author}]{hyperref}
\begin{document}
\title{Realizing a compact, high-fidelity, telecom-wavelength source of multipartite entangled photons}
\author{Laura dos Santos Martins}
    \email[Correspondence email address: ]{laura.dos-santos-martins@lip6.fr}
    \affiliation{Sorbonne Université, CNRS, LIP6, 4 Place Jussieu, Paris F-75005, France}

\author{Nicolas Laurent-Puig}
    \affiliation{Sorbonne Université, CNRS, LIP6, 4 Place Jussieu, Paris F-75005, France}

\author{Pascal Lefebvre}
    \affiliation{Sorbonne Université, CNRS, LIP6, 4 Place Jussieu, Paris F-75005, France}

\author{Simon Neves} 
\affiliation{Sorbonne Université, CNRS, LIP6, 4 Place Jussieu, Paris F-75005, France}
    \affiliation{Department of Applied Physics, University of Geneva, CH-1211 Genève, Switzerland}
   
\author{Eleni Diamanti}
    \affiliation{Sorbonne Université, CNRS, LIP6, 4 Place Jussieu, Paris F-75005, France}

\date{\today}

\begin{abstract}
Multipartite entangled states are an essential building block for advanced quantum networking applications. Realizing such tasks in practice puts stringent requirements on the characteristics of the states in terms of fidelity and generation rate, along with a desired compatibility with telecommunication network deployment. Here, we demonstrate a photonic platform design capable of producing high-fidelity Greenberger–Horne–Zeilinger (GHZ) states, at telecom wavelength and in a compact and scalable configuration. Our source  relies on spontaneous parametric down-conversion in a layered Sagnac interferometer, which only requires a single nonlinear crystal. This enables the generation of highly indistinguishable photon pairs, leading by entanglement fusion to four-qubit polarization-entangled GHZ states with fidelity up to $(94.73 \pm 0.21)\%$ with respect to the ideal state, at a rate of $1.7$~Hz. We provide a complete characterization of our source and highlight its suitability for practical quantum network applications.

\end{abstract}

\maketitle

\section{Introduction}
\label{sec:introduction}
In recent years, significant efforts have been made towards establishing quantum information networks able to interconnect remote quantum systems, giving rise to applications ranging from information-theoretically secure communication to distributed quantum computing and sensing~\cite{QNetwork}. Photonic resources, despite being inherently probabilistic due to the non-interactive nature of photons, are the workhorse of quantum networks thanks to their robustness to noise and ability to transfer information over long distances. Among the photonic platforms developed to match the requirements of quantum information applications, the emission of photon pairs by spontaneous parametric down-conversion (SPDC) in nonlinear crystals has emerged as the most natural way to generate entangled photonic qubits, with information often encoded in the polarization degree of freedom. In particular, sources based on Sagnac interferometers that exhibit a remarkable intrinsic stability have demonstrated their ability to generate entangled Bell states of excellent quality~\cite{Sagnac2004Shi,Sagnac2006Kim,fedrizzi2007Sagnac,Sagnac2008Kuzucu,Sagnac2012Pulsed,Sagnac2015CW,vergyris2017Sagnac}. This has enabled the implementation of a great variety of entanglement-based quantum communication and cryptographic protocols, including in the highly demanding setting of satellite-to-ground links~ \cite{SagnacSatelliteQKDJWP,pelet2022signatures,pavelQChannel2023,nevesQChannel2023}.

Despite this significant progress, a much richer range of quantum networking functionalities can be reached when moving beyond two-party communication tasks, which involves the use of multipartite entangled states, such as, for instance, the Greenberger–Horne–Zeilinger (GHZ) states. These functionalities include secret sharing, conference key agreement, anonymous communication, electronic voting, or networks of sensors, among others, where in all cases the use of the entangled resource provides a provable advantage in terms of data security or privacy~\cite{SagnacSecretSharingRarity,SagnacMultipartyStockholm,unni2019anonymous,ConferenceKeyAgree,centrone2022voting,shettell2022sensornetworks}. In practice, multiple independent photon pair sources can be `fused' into multipartite entangled states. Remarkably, this approach has been used to demonstrate entanglement between 12 photons by combining 6 independent SPDC sources~\cite{12GHZ}. Nevertheless, from an application perspective the use of such states has consisted in general in proof-of-principle lab-scale protocol demonstrations, incompatible with the requirements of large-scale quantum networks. Major challenges include the extraordinarily high level of indistinguishability required between photons involved in the entanglement fusion operation, while the size and stability of the optical setup are crucial for developing sources suitable for deployment outside the laboratory. Additionally, the achieved fidelity is a critical parameter for reaching operation regimes where a quantum advantage can be demonstrated in realistic scenarios~\cite{verification2016}. 

In this work, we realize a compact source of high-quality GHZ-states at telecom wavelength, leveraging an original design based on spatial multiplexing in a Sagnac configuration. In particular, two sources are vertically stacked inside the same periodically-poled KTP (ppKTP) crystal and Sagnac interferometer, resulting in a layered Sagnac source~\cite{nevesMultiplexing}, able to distribute two independent photon pairs with a single optical setup. Similar methods have been very recently applied to generate hyperentanglement~\cite{zhao2024direct} and to enhance teleportation~\cite{daurelio2024boosted}. In this configuration, photons are optimally coupled into single-mode fibers (SMF) due to collinear emission, contrary to multiplexed sources based on ring distribution of photons emitted in BBO crystals~\cite{smaniaMultiplexing,guilbertMultiplexing}. Independent photon pairs from two different layers are combined into a 4-photon entangled state via entanglement fusion. The indistinguishability between independent photons is optimized as they are emitted in the same crystal and follow a similar path, so the resulting states naturally display a high fidelity with respect to the target GHZ states. As this design can be used in principle to stack a larger number of photon pair sources, it opens perspectives for the generation of large GHZ states maintaining high fidelity, in a stable and compact setup, suitable for deployment in a large-scale quantum network.

\section{Experimental Setup}
\label{sec:exp_setup}

\begin{figure*}[!t]
\centering
 \includegraphics[width=\textwidth]{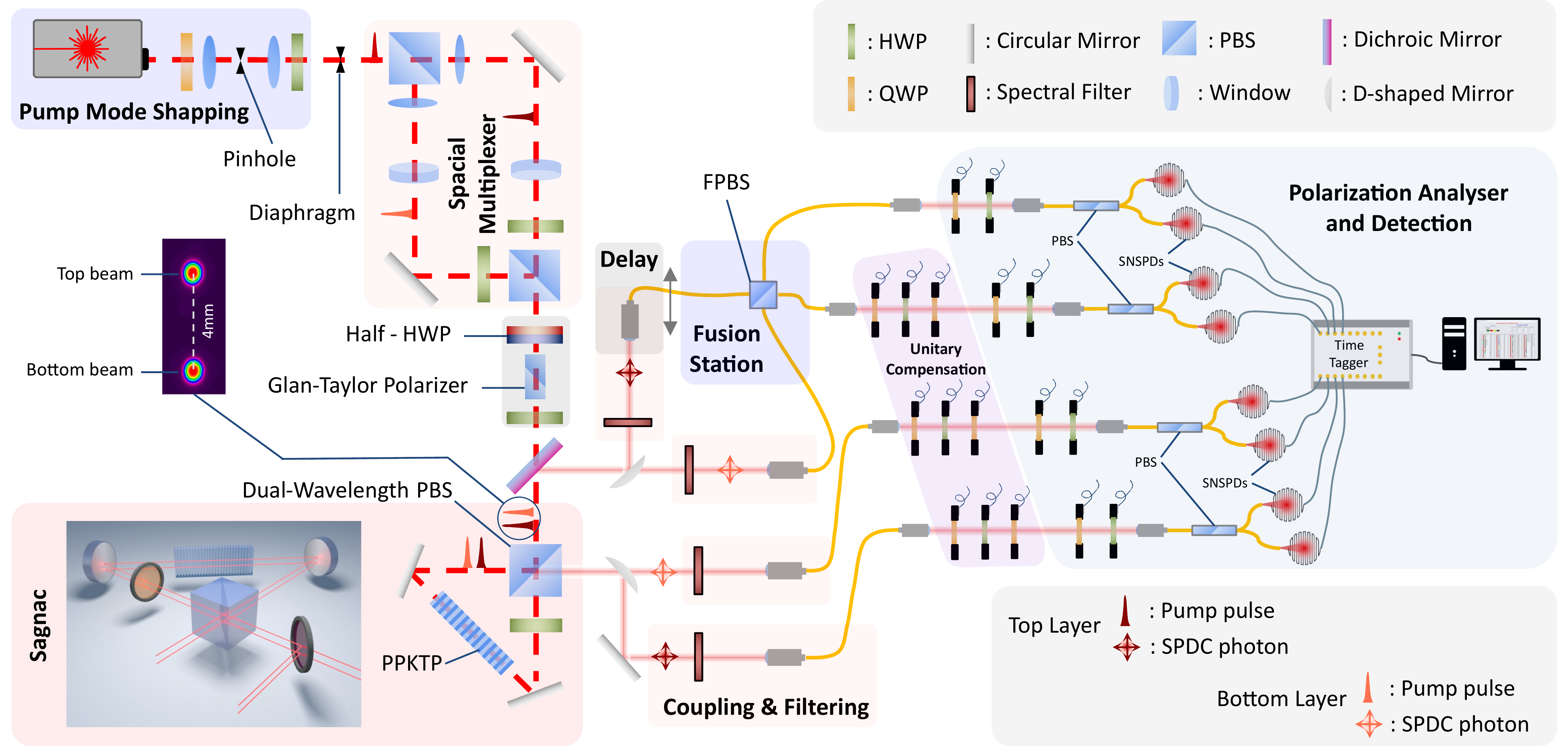}
        \caption{Layered Sagnac GHZ source. \textbf{Laser pump:} A Ti:Sapphire laser (Coherent Mira-HP) with an average power of 3.4~W emits 2~ps pulses at a wavelength of 775~nm with a repetition rate of 76~MHz. \textbf{Spatial mode shaping:} The spatial mode of the laser is shaped into a Gaussian profile. \textbf{Spatial multiplexer:} The pump pulses are split into two parallel beams: the top and bottom layers, horizontally and vertically polarized, respectively. \textbf{Polarization shaping:} Both layers are diagonally polarized, in order to maximize the fidelity of output states with respect to the Bell state. \textbf{Sagnac interferometer:} Photon pairs are probabilistically generated via type-II SPDC in a ppKTP crystal (30mm-long, 46.2~$\mu$m poling period, provided by Raicol) and entangled in polarization in the Sagnac loop, resulting in the output state $(\ket{H}_s\ket{V}_i+e^{i\theta}\ket{V}_s\ket{H}_i)/\sqrt{2}$. \textbf{Coupling and Filtering:} After filtering the single photons with a dichroic mirror, 1100~nm long pass filters and 1.3~nm ultra-narrowband filters, the bottom layer photons are reflected on half-circle shaped mirrors, while the top layer photons are transmitted over them. The photons are coupled to SM fibers with 12~mm focal lens. \textbf{Fusion station:} The mechanical delay on the bottom layer idler photon is fine tuned such that both idler photons (from the top and bottom layers) arrive simultaneously to the FPBS. If each of them is transmitted to different outputs of the FPBS, and conditioned on fourfold coincidences, a GHZ state $(\ket{HHHH}+e^{i\delta}\ket{VVVV})/\sqrt{2}$ is generated. \textbf{Unitary compensation:} Three sets of QWP-HWP-QWP rotate the final state to the $\ket{GHZ}_{\delta=0}$ state. \textbf{Polarization Analyser and Detection:} A set of HWP-QWP-FPBS is used to map each photon's polarization to the spatial degree of freedom. The fibers are connected to superconducting nanowire single-photon detectors (ID281 SNSPD) that, in turn, are linked to a time tagger that allows to count and correlate detection events for the analysis.}
        \label{fig:source_setup}
\end{figure*}

Our goal is to generate 4-photon polarization-entangled GHZ states, taking the form $\ket{GHZ}_\delta=(\ket{HHHH}+e^{i\delta}\ket{VVVV})/\sqrt{2}$, where $\delta$ is the phase between the two components of the state. The key concept behind our experimental setup consists in using a single, layered Sagnac interferometer and a single nonlinear crystal to generate two entangled-photon pairs, with light coming from a spatially-multiplexed pump laser. This makes the source particularly compact and allows to generate highly indistinguishable pairs, which are then fused with an entanglement fusion gate before being measured. The setup is displayed in Fig.~\ref{fig:source_setup}.

\subsection{Multiplexed Pump}
We pump the crystal with a $\SI{3.4}{\watt}$-average power mode-locked Titanium-Sapphire Laser. The central wavelength is approximately $\SI{775}{\nano\meter}$, and the pulse duration is $\SI{2}{\pico\second}$. Laser pulses are emitted at a $\SI{76}{\mega\hertz}$ frequency. We first shape the spatial mode of the pump laser by placing a pinhole in the focal plane of two $\SI{50}{\milli\meter}$-focal length lenses. The lower intensity rings of the resulting Airy disk are filtered out with a diaphragm, so we are left with a close-to-Gaussian beam profile. This later facilitates the coupling of photons into SMFs.

We then send the pump beam to a spatial multiplexer, which allows to pump the same Sagnac interferometer and ppKTP crystal with two parallel beams. The pump is split into two paths with a polarizing beam-splitter (PBS), and the  transmission/reflection (T/R) ratio is tuned with a half-wave plate (HWP) and a quarter-wave plate (QWP) placed before the PBS. We thus tune the emission probability of SPDC pairs in each layer, to later compensate for eventual coupling and loss unbalance. The transmitted (reflected) path after the PBS features a tilted optical window that deviates the beam a few millimeters down (up), followed by a HWP that guarantees the total transmission (reflection) of the beam on the succeeding PBS. Finally, the output of the spatial multiplexer is composed of two parallel beams, the top and bottom layers, horizontally and vertically polarized, respectively, approximately separated by 4 mm, as shown in Fig.~\ref{fig:source_setup}.

The last stage of the multiplexed-pump shaping is to make sure both beams have the same polarization, so that we can avoid fine tuning the polarization of each layer individually. To achieve this, we first set them up, roughly, to the horizontal polarization with a custom HWP - the top (bottom) half of the plate is oriented at $\SI{0}{\degree}$ ($\SI{45}{\degree}$) to the horizontal direction. We then project both beams on the exact same linear polarization, using a Glan-Taylor polarizer, and with minimal losses. A HWP at $\simeq\SI{22.5}{\degree}$ finally sets up both layers to an approximately diagonal polarization. Fine-tuning of this HWP allows us to simultaneously optimize the fidelity of the states produced in both layers, with respect to the Bell state.

\subsection{Layered Sagnac}
The next step to generate a 4-photon GHZ state is the simultaneous emission of two independent Bell pairs. For that purpose, we send the multiplexed pump to a Sagnac interferometer, featuring a nonlinear ppKTP crystal, which is optimized for the emission of quasi-spectrally-separable photons at a $\SI{1550}{\nano\meter}$-wavelength, by type-II SPDC. The pump horizontal and vertical components are spatially split by a dual-wavelength PBS, traveling through the Sagnac loop in clockwise or counterclockwise direction, respectively. The vertically-polarized pump component is down-converted into a signal-idler pair in the polarization state $\ket{H}_s\ket{V}_i$. The horizontally-polarized clockwise-propagating component is flipped to vertical polarization by a broadband HWP, before undergoing the same down-conversion. In the counterclockwise direction, the SPDC photons are flipped into $\ket{V}_s\ket{H}_i$ by the same broadband HWP. The photon pairs are then spatially separated by the dual-wavelength PBS. At the output of the Sagnac loop, the overlap of the beams makes it impossible to know whether the pair was emitted in the clockwise or anticlockwise direction, resulting in a superposition of both possibilities. This corresponds to the Bell state $\ket{\psi}_\theta = (\ket{H}_s\ket{V}_i+e^{i\theta}\ket{V}_s\ket{H}_i)/\sqrt{2}$ , where $\theta$ is the phase between the two propagation paths. Additionally, the crystal is $\SI{10}{\milli\meter}$-large, and can thus receive the two parallel beams from the multiplexed pump. Provided the pump undergoes minimal spatial distortions, the resulting Sagnac interferometer then features two layers with the exact same optical properties, from which we collect two independent and indistinguishable Bell pairs. At the output of the layered Sagnac source, the pump is filtered out with a dichroic mirror and a 1100~nm long pass filter. Finally, the photons from two different layers are spatially separated by half-circle shaped mirrors, so photons from the bottom layer are reflected, and those from the top layer are transmitted.

\subsection{Entanglement Fusion}
The final step to generate the multipartite state is the fusion of two entangled pairs from different layers of the Sagnac source. We take the idler photon from each Bell pair, coupled to SMFs, and make them interfere on a fiber-PBS (FPBS). We use a motorized delay stage to fine tune the temporal overlap of the interfering photons (see Fig.~\ref{fig:hom} in the Appendix). We post-select on the events where each photon exits at a different port of the FPBS. Consequently, if they are indistinguishable, it is impossible to distinguish if both photons were transmitted or reflected, i.e., if both photons are horizontally or vertically polarized. In other words, conditioned on fourfold coincidences, the 4-photon state is projected onto a superposition of both situations, which takes the form $\ket{GHZ}_\delta=(\ket{HHHH}+e^{i\delta}\ket{VVVV})/\sqrt{2}$ up to local unitaries carried out by the single-mode fibers, where $\delta$ is defined by the $\theta$ of each Bell pair. In order to increase the fidelity with respect to this target state, we employ 1.3~nm ultra-narrowband filters on the path of all four photons, which increases the spectral purity of the Bell pairs (see section~\ref{sec:jsi}).

\subsection{Measurement}
Finally, a set of HWP-QWP-FPBS is used to map the photon's polarization to the spatial degree of freedom - each output fiber of the FPBS will only transmit the photons that collapsed to a specific polarization eigenstate, one to $\ket{H}$ and the other to $\ket{V}$. The fibers are connected to superconducting nanowire single-photon detectors (SNSPD), whose counts are recorded by a time tagger for the analysis of the detection events. By carefully configuring the coincidence windows and time delays between the different detection channels, we effectively mitigate noise while capturing the fourfold event instances.

We also remark that in many quantum information applications it is essential to ensure a specific form for the states being produced. More specifically, regarding the GHZ states generated with our source, we need to address the transformations induced by the single-mode fibers and apply the corresponding inverse transformations to recover a true GHZ state, expressed as $U^{\dagger}_{\text{fiber}}(\ket{GHZ}_{\delta} \bra{GHZ}_{\delta}) U_{\text{fiber}} = \ket{GHZ}_{0} \bra{GHZ}_{0}$, where $U_{\text{fiber}}$ is a tensor product of local unitaries. This is realized with three sets of QWP-HWP-QWP, one in each path of the four entangled photons (see section \ref{sec:app:uc}). To find the configuration of these waveplates, we measure the density matrix of the state produced with our source via quantum state tomography (QST) (see section \ref{sec:QST} for details), and run an optimization method to find the three-qubit unitary that rotates the state to its desired form.
\section{Results}
\label{sec:results}

\begin{figure}[b!]
\centering
\includegraphics[width=\columnwidth]{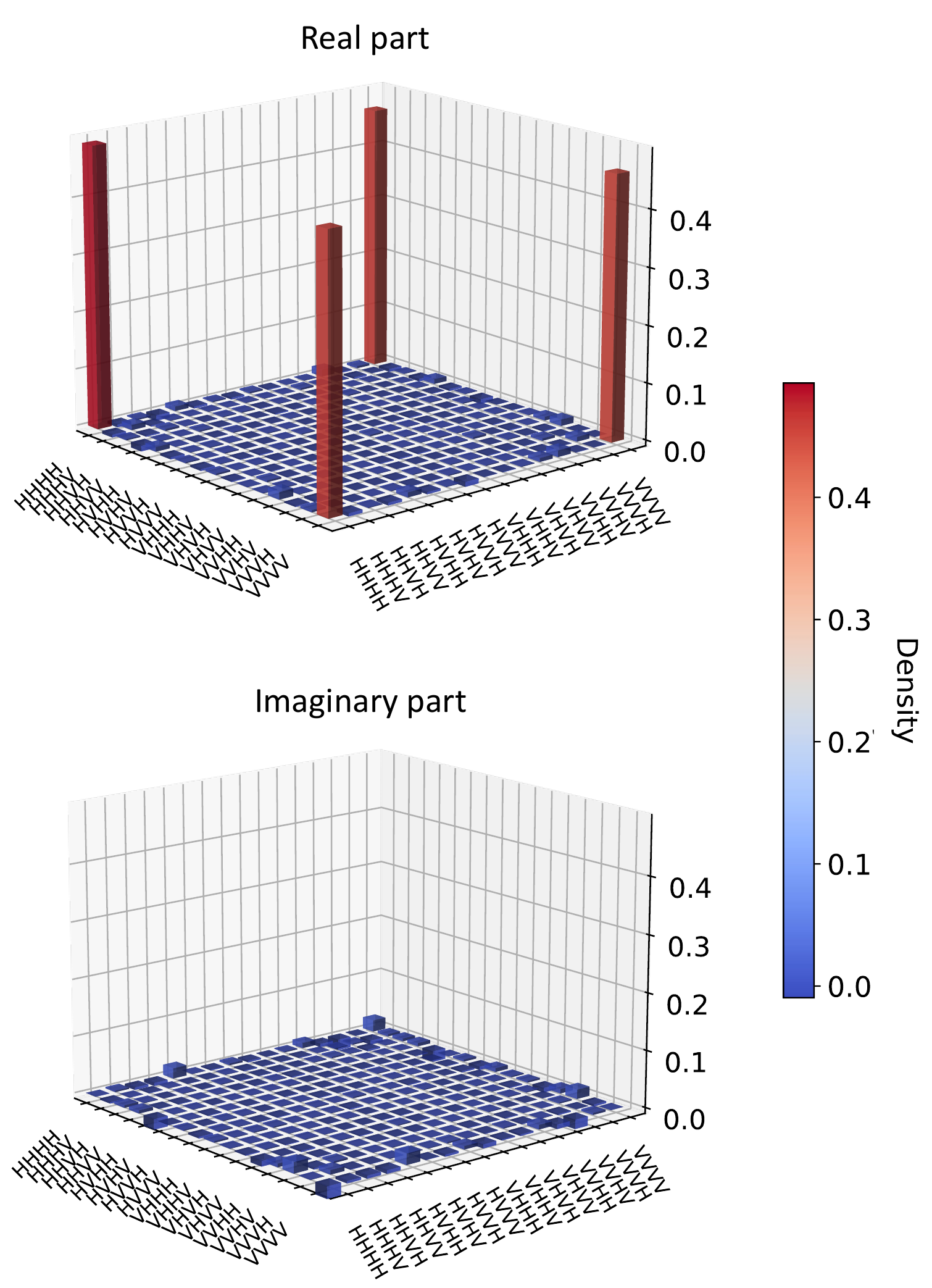}
 \caption[dm]{Experimental density matrix of the state produced with our source, estimated via quantum state tomography. The pump power was set to 63 mW (57 mW) in the top (bottom) layer, yielding a rate of 1.7 GHZ states per second for an acquisition time of 600~s per basis. The fidelity to the GHZ state is $\mathscr{F}=(94.73 \pm 0.21)\%$.}
  \label{fig:dm}
\end{figure}

By pumping the crystal with 62 mW and 57 mW, in the top and bottom layers, respectively, we generate an average of 1.7 fourfold coincidences per second. We characterize our source by quantum state tomography, using 97 measurement bases - 81 possible combinations of the X, Y and Z basis for each photon plus 16 possible combinations of Z and -Z to estimate the relative efficiencies of different outcomes. We use the linear regression estimation method for fast quantum state reconstruction (see details in section \ref{sec:QST}). The resulting estimated state $\rho_{E}$ is displayed in Fig.~\ref{fig:dm}, and we measure a fidelity $\mathscr{F}=\vert\bra{GHZ}\rho_{\text{exp}}\ket{GHZ}\vert^2 = (94.73 \pm 0.21)\%$ with respect to the ideal GHZ state.

Furthermore, we analysed the state fidelity obtained with our layered-Sagnac source as a function of the generation rate, determined by the pump power. Besides the four-fold events, we additionally recorded the five-fold events and reconstructed the density matrix of each state, either by simply considering the four-fold events or by subtracting the five-fold from the desired four-fold, i.e., by correcting for the events that had two emitted pairs in the same layer, which we refer to as high-order emission. Even higher-order detection events were considered negligible.

\begin{figure}[!htbp]
\centering
\includegraphics[width=\columnwidth]{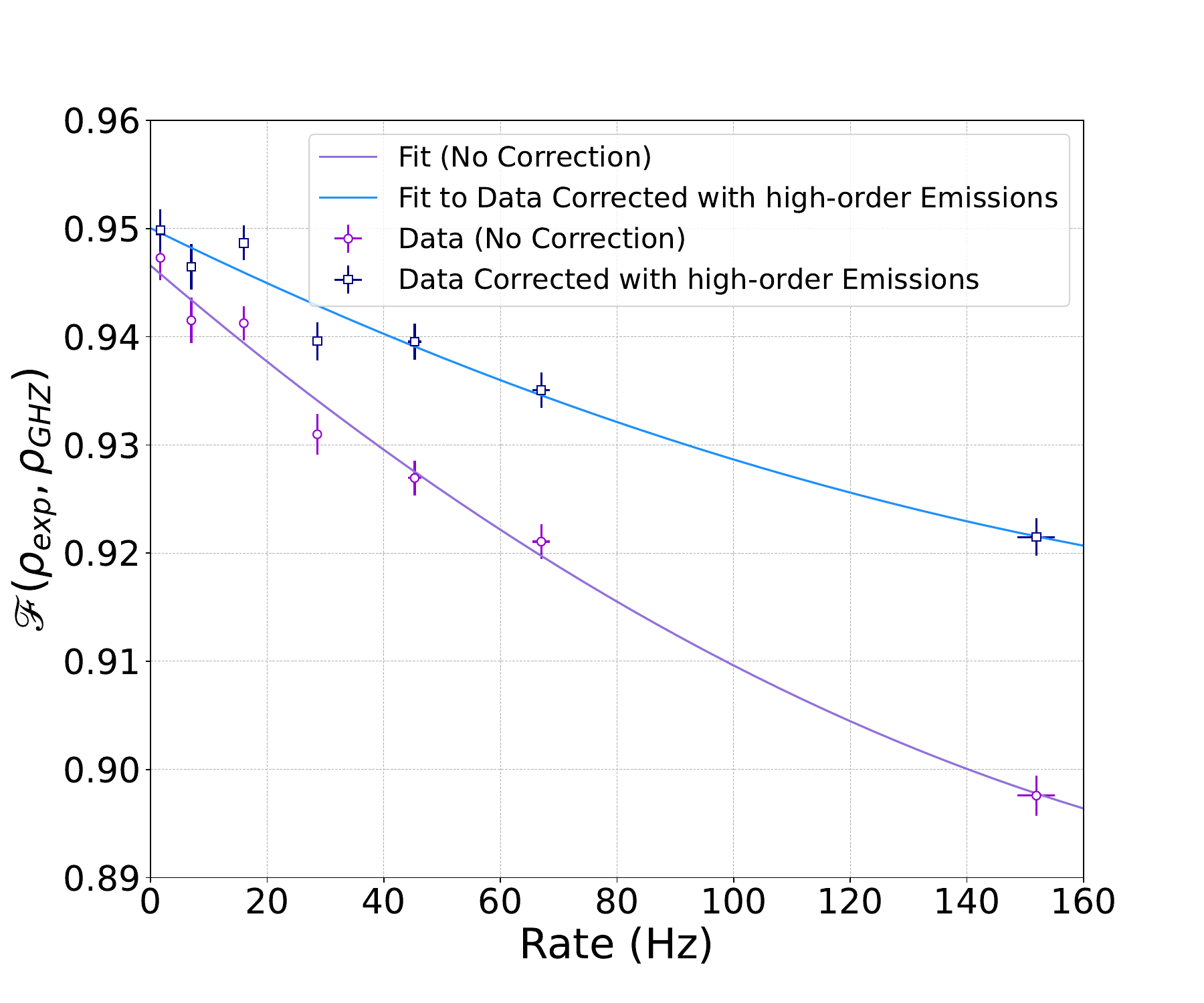}
 \caption[dm]{Experimental state fidelity with respect to the GHZ state, $\mathscr{F}(\rho_{\text{exp}},\rho_{GHZ})$, as a function of the four-fold coincidences rate, for density matrices directly estimated with all the four-fold events detected (purple) and with the counts corrected to exclude the high-order emission events (blue). We observe that the two curves meet at a zero rate as expected within the error bars. The lack of perfect convergence is due to the residual imperfections in the higher-order emission correction.}
  \label{fig:fidelity_vs_rate}
\end{figure}

The maximum rate we can achieve at maximum pump power is around 152~Hz for a fidelity of $(89.71 \pm 0.19)\%$, or $(92.14 \pm 0.17)\%$ with the high-order emission correction. It is clear that as we increase the GHZ state generation rate, the difference between the estimated fidelity with the corrected and non-corrected data becomes more significant. The reason for this is the fact that the ratio between high-order emissions and the desired single-pair per layer events is  $\sim \frac{p^3}{p^2}=p$, where $p$ is the probability associated with one SPDC process. This means that, as the rate $\propto p^2$ increases, the high-order emission events become more predominant and, therefore, their correction has a bigger impact on the estimated fidelity. However, since we can only subtract the five-fold events that were detected and some of the high-order emission events will not be recorded as such, due to losses and inefficiencies in the setup, we can never fully compensate for this effect, thus still observing a decreasing fidelity for an increasing rate. Finally, we believe that the fidelity observed at very low power is mostly limited by the extinction ratio of the FPBS and some residual spectral correlations.
\section{Conclusions and Perspectives}
\label{sec:conclusions}
In this work, we have successfully demonstrated an original design of a four-qubit GHZ state source that combines several attractive features. Based on a layered Sagnac configuration and exploiting spatial multiplexing, our source is exceptionally compact and stable since a single optical setup is used for generating the two photon pairs necessary to produce the multipartite entangled state. This setup ensures that, beyond filtering the spectral correlations between individual SPDC pairs, the photons' indistinguishability is guaranteed, hence leading to high-fidelity GHZ states by design. Additionally, considering the collinear emission of photon pairs, we can conclude that this design is optimized such that we can obtain the best fidelity to rate compromise, making it suitable for implementing quantum communication protocols, within reasonable time frames, while having access to high-fidelity states. Furthermore, our source operates at telecom wavelength, ensuring compatibility with deployment in standard network infrastructures.

\begin{table}[h!]
\centering
\begin{tabular}{c|c|c|c}
\textbf{Ref.} & \textbf{$\mathscr{F}_{max}$ [\%]} & \textbf{Rate [Hz]} & \textbf{$\lambda$ [nm]} \\
\hline
This work & 94.7 & 1.7 & 1550 \\
\hline
Jonathan W. \textit{et al.} \cite{Webb:24} & 93.3 & 5.5 \footnotemark[1]{}\footnotetext[1]{We deduced this value using the information provided in the corresponding publication.} & 1500 \\
\hline
D. Wu \textit{et al.} \cite{PhysRevLett.127.230503} & 95.7 & 0.4\footnotemark[1]{} & 1500 \\
\hline
 C. Zhang \textit{et al.} \cite{Zhang:16} & 97.9 & 0.4 & 780\\
\hline
X.-L. Wan \textit{et al.} \cite{PhysRevLett.117.210502} & 83.3 & 6000 & 788 \\
\hline
\end{tabular}
\caption{State-of-the-art for 4-photon GHZ state sources.}
\label{tab:spdc_cources}
\end{table}

To assess the performance of our source with respect to previous demonstrations, we summarize a few notable, state-of-the-art 4-photon GHZ state results in Table~\ref{tab:spdc_cources}. At telecom wavelengths, the fidelity-rate results reported by Jonathan W. \textit{et al.}~\cite{Webb:24} and D. Wu \textit{et al.}~\cite{PhysRevLett.127.230503} are comparable to our findings. More specifically, based on the fit shown in Fig.~\ref{fig:fidelity_vs_rate}, we estimate that our source would yield a fidelity with respect to the GHZ state of $\mathscr{F}=94.4\%$ and $\mathscr{F}=94.5\%$ at the respective rates reported in these references. Furthermore, it is worth noting that in the work done by Jonathan W. \textit{et al.}, an aperiodically-poled KTP crystal is used, which supports the idea of incorporating a similar crystal in a layered Sagnac setup as a way to boost its performance, as we also discuss below. Additionally, while the sources reported by C. Zhang \textit{et al.}~\cite{Zhang:16} and X.-L. Wan \textit{et al.}~\cite{PhysRevLett.117.210502} operate at wavelengths not suited for current telecommunication infrastructures, they still stand out for their high fidelity and rate, respectively, providing good platforms for proof-of-principle experiments. Considering the aforementioned, it is clear that our source is on par with the state-of-the-art photonic four-partite entanglement devices while also offering the  advantages discussed previously. Please note that in the above discussion we have omitted error bars in the fidelities for simplicity. Furthermore, regarding the work done by D. Wu \textit{et al.}, we stress that the estimation of the fidelity is based on an entanglement witness measurement rather than full quantum state tomography.

A limitation of our present setup comes from the high-order photon pair emission, which is intrinsic to all SPDC-based sources. One way to mitigate this is by reducing the pump power, although this naturally comes at the expense of decreasing the state generation rate. The spectral purity is another aspect still contributing to the noise, manifested in the interference visibility of $90.6\%$ shown in section~\ref{sec:HongOuMandel}. Although this quantity is limited to approximately 95\%, given the time jitter of the interfering photons compared to their coherence length~\cite{photon_emission_jitter}, one could improve the visibility by further filtering the photon pairs. 
However, as it can be seen in Fig.~\ref{fig:jsi_results} there is a slight overlap between the central mode we want to preserve and the side modes we want to filter out. This develops, once more, into a fidelity vs rate trade-off. Alternatively, a promising path as mentioned above is to use an aperiodically-poled KTP crystal~\cite{PhysRevA.98.053811}, designed to generate intrinsically spectrally-pure photon pairs. This would remove losses caused by the ultra-narrow filter altogether. Finally, the limited extinction ratios of the FPBS ranging from 25~dB to 33~dB in our setup could be improved.

In terms of scalability, we remark that it is in principle possible to scale the setup to larger states without building extra Sagnac interferometers, as long as larger crystals can be produced to fit several parallel beams in. The rate of course will still be limited by the probabilistic nature of SPDC emissions. Additionally, as we increase the number of simultaneous SPDC pairs, the alignment, and in particular the coupling, will eventually become challenging. Still, even stacking a few sources can save up considerable resources, which can be an important asset in the emerging large-scale quantum communication networks.

In summary, our results illustrate the feasibility of an original telecom GHZ state source design, which combines compactness and stability with the capability to produce high-fidelity states at reasonable rates, paving the way to the demonstration of compelling multipartite quantum information tasks in real-world network settings, outside research laboratories. 
\section{Acknowledgments}
\label{sec:acknowledgements}
We acknowledge financial support from the European Union’s Horizon 2020 framework programme under the Marie Sklodowska Curie innovation training network project AppQInfo, Grant No. 956071 (LdSM), the Horizon Europe research and innovation programme under the project QSNP, Grant No. 101114043 (ED), the European Research Council Starting Grant QUSCO, Grant No. 758911 (ED, SN, NLP), and the PEPR integrated project QCommTestbed, ANR-22-PETQ-0011, part of Plan France 2030 (ED, PL). 
\bibliographystyle{ieeetr}
\bibliography{references.bib}
\clearpage
\onecolumngrid
\appendix*
\section*{Supplementary Material}
\label{sec:appendix}
\subsection{Quantum State Tomography and Errors}
\label{sec:QST}
We conduct quantum state tomography using linear regression estimation \cite{LR-QST} and fast maximum likelihood estimation \cite{PhysRevLett.108.070502}. Additionally, we correct the photon counts with the relative efficiencies of each photon's path (from the coupling to the detectors), relying on the $\{Z, -Z\}$ measurement results of each qubit. This approach allows us to reconstruct the density matrix description of the measured state and determine its fidelity with respect to the target state. To assess the uncertainties associated with the reconstructed states, we employ the Monte Carlo method by sampling from Poissonian photon counting statistics and Gaussian QHP-HWP rotation angle distributions (which incorporates the systematic measurement basis error). For each density matrix estimation, we simulate 500 new data samples incorporating the respective uncertainty distributions, reconstructing new density matrices to compute the average fidelity and standard deviation. Furthermore, the fidelity may be affected by slow thermal fluctuations due to the extended duration of the measurements, particularly for the state estimations at low rates.

\subsection{Bell State Fidelity}
\label{sec:BellState}
The four-partite entanglement results shown in section \ref{sec:results} were obtained based on the fusion of two Bell pairs, each produced in a different spatial layer of the setup. While pumping the crystal with 2 ps pulses and with and average power of 500 mW for each layer, we measured a quantum state fidelity of $(98.20 \pm 0.01)\%$ and $(97.56 \pm 0.01)\%$, respectively, with respect to the maximally entangled state $\phi=(\ket{HH}+\ket{VV})/\sqrt{2}$. The associated detection rate was around 250 kHz for each pair, in the absence of the ultra-narrowband filters. The results were estimated after applying the double emission correction.

\subsection{Joint Spectral Intensity of SPDC Photon Pairs}
\label{sec:jsi}
In our setup, photon pairs are generated in a ppKTP crystal under the conservation of energy and the quasi-phase matching conditions. These conditions govern the strength of the interaction between the electromagnetic field of the pump and the crystal. This interaction results in the production of photon pairs that are not spectrally pure. In other words, if we measure the frequency of one, we can get information about the frequency of the other. This becomes a problem once we interfere two photons from different pairs. Intuitively, if we consider the entanglement fusion used to generate the GHZ state (see Fig.~\ref{fig:source_setup}), this could possibly mean that by measuring the photons' frequencies after the FPBS, one could distinguish the different pairs, which implies that the visibility of the interference is compromised. For this reason, if we want to maximize the state fidelity with respect to the GHZ state, we need to filter the photon pairs, such that they are no longer correlated in spectrum. In order to determine the appropriate bandwidth and the central wavelength for such filtering, it is necessary to measure the joint spectral intensity (JSI) of the SPDC pairs.

\begin{figure}[H]
\centering
\subfloat[]{\includegraphics[width=0.44\columnwidth]{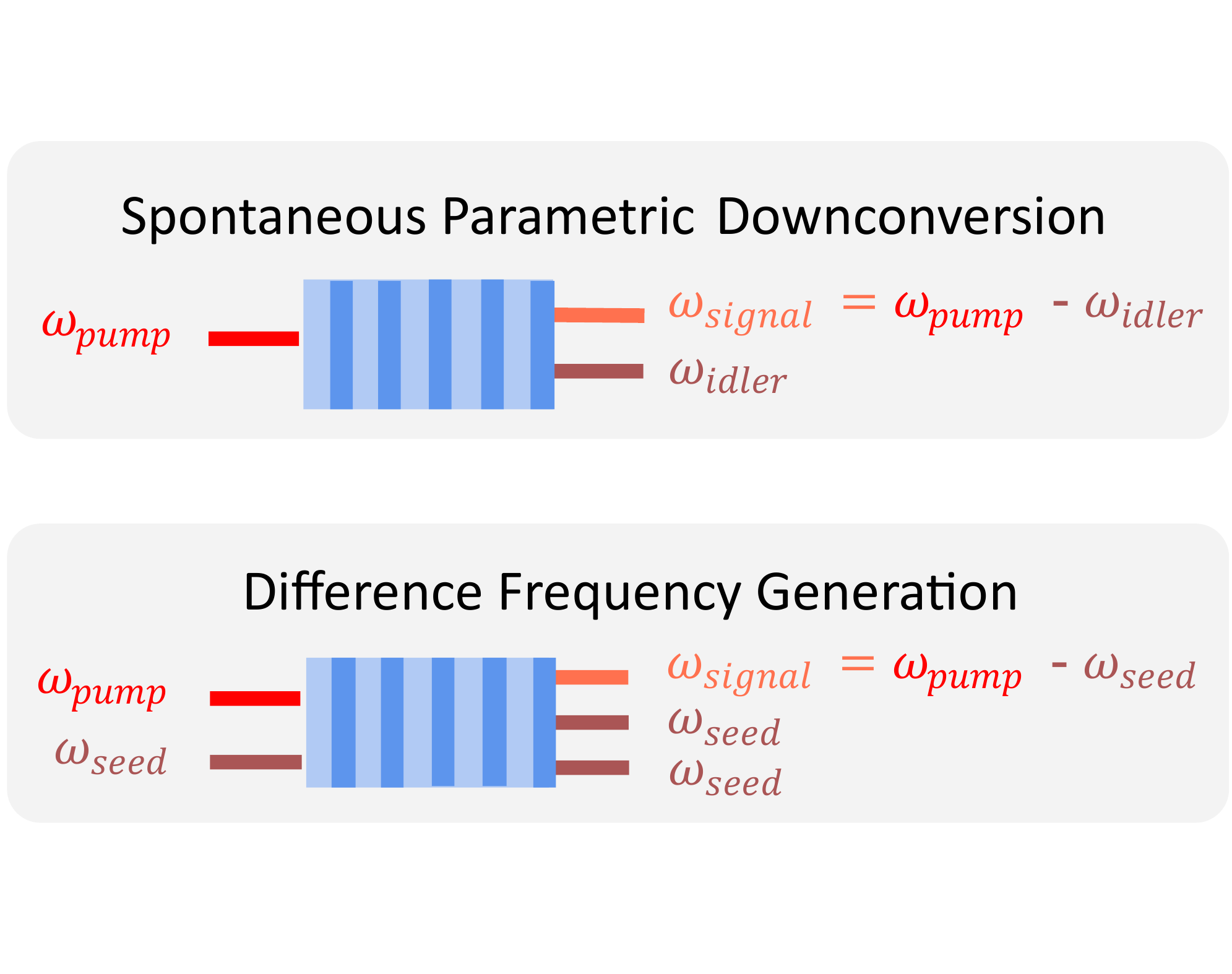}}
\subfloat[]{\includegraphics[width=0.44\columnwidth]{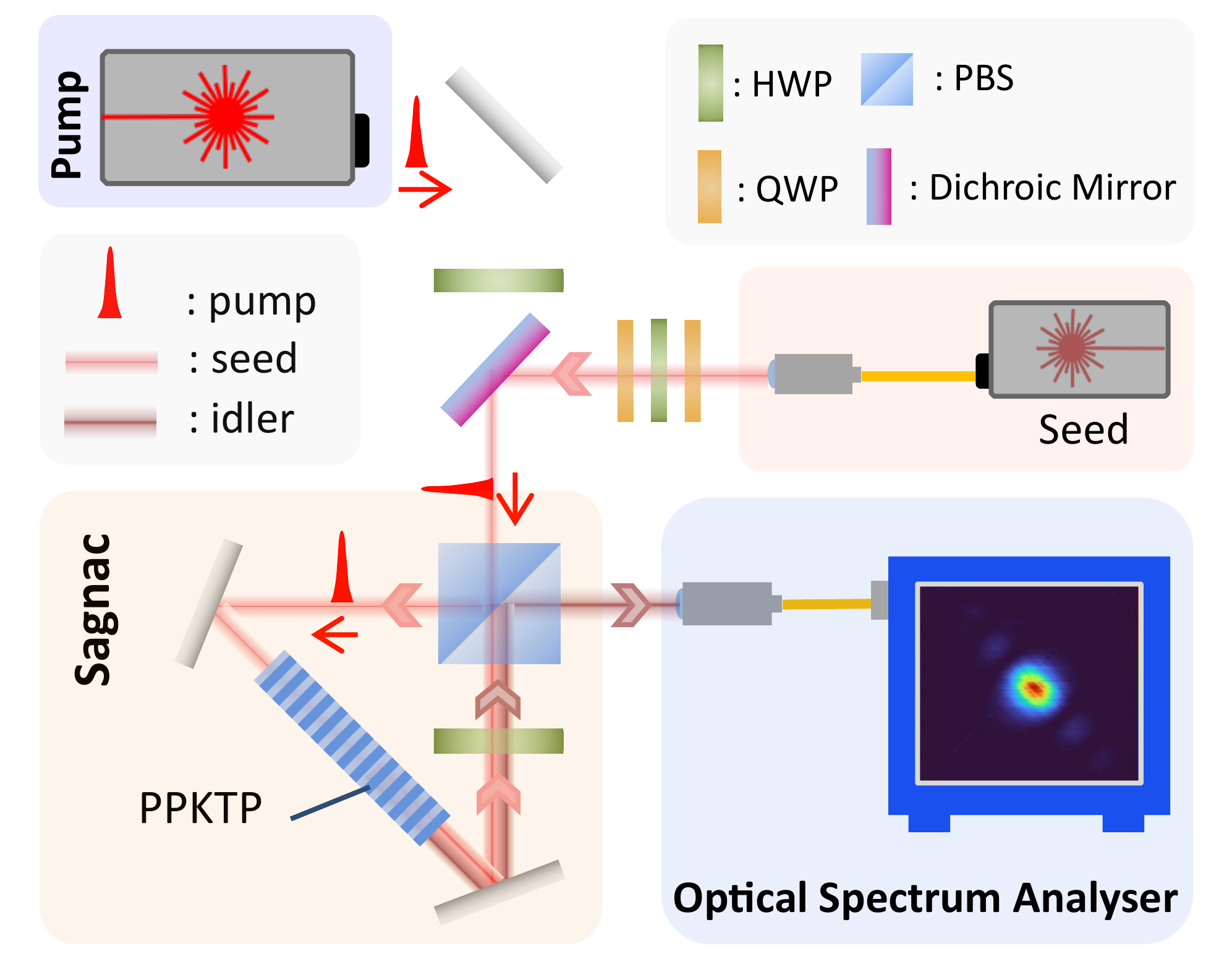}}
\caption{a) Schematic for (top) SPDC - a pump photon with frequency $\omega_p$ is converted into two photons of lower energy: $\omega_s$ and $\omega_i=\omega_p-\omega_s$, respectively; (bottom) DFG - once the pump $\omega_p$ exits the crystal, the seed, $\omega_s$, stimulates the emission of two photons: one of the same frequency $\omega_s$ and another with the remaining energy, $\omega_i=\omega_p-\omega_s$. b) Joint Spectral Intensity measurement setup. The pump - previously described - and seed - CW at telecom band (tunable) with 20 pm linewidth - lasers are sent to the Sagnac loop polarized vertically. The interaction of the two electromagnetic fields with the ppKTP crystal results in difference frequency generation, whose outcome is composed of two frequencies: the frequency of the seed (vertically polarized), $\omega_s$, and the difference between the pump and the seed frequencies, $\omega_i=\omega_p-\omega_s$, horizontally polarized. The HWP rotated to 45 degrees, together with the PBS, filter out the seed from the idler, reflecting the latter to be later collected by the optical spectrum analyser (EXFO OSA20).}
\label{fig:jsi}
\end{figure}

We experimentally measured the spectrum of the photon pairs produced in our source, relying on a stimulated emission method~\cite{doi:10.1080/09500340.2018.1437228}. More specifically, we use difference frequency generation (DFG), which can be seen as the classical analog of SPDC (see Fig.~\ref{fig:jsi} a)), to boost the emission of pairs, such that they can be measured by a low noise optical spectrum analyzer (OSA). This is only possible since SPDC and DFG share the same phase matching condition and therefore, if we use the same pump spectrum, we will measure the same spectral correlation function~\cite{PhysRevLett.111.193602}. The advantage of this method is that it generally allows for high resolution measurements, since it is only limited by the frequency steps of the laser and the resolution of the OSA.

\begin{figure*}[!htb]
\centering
\includegraphics[width=1\textwidth]{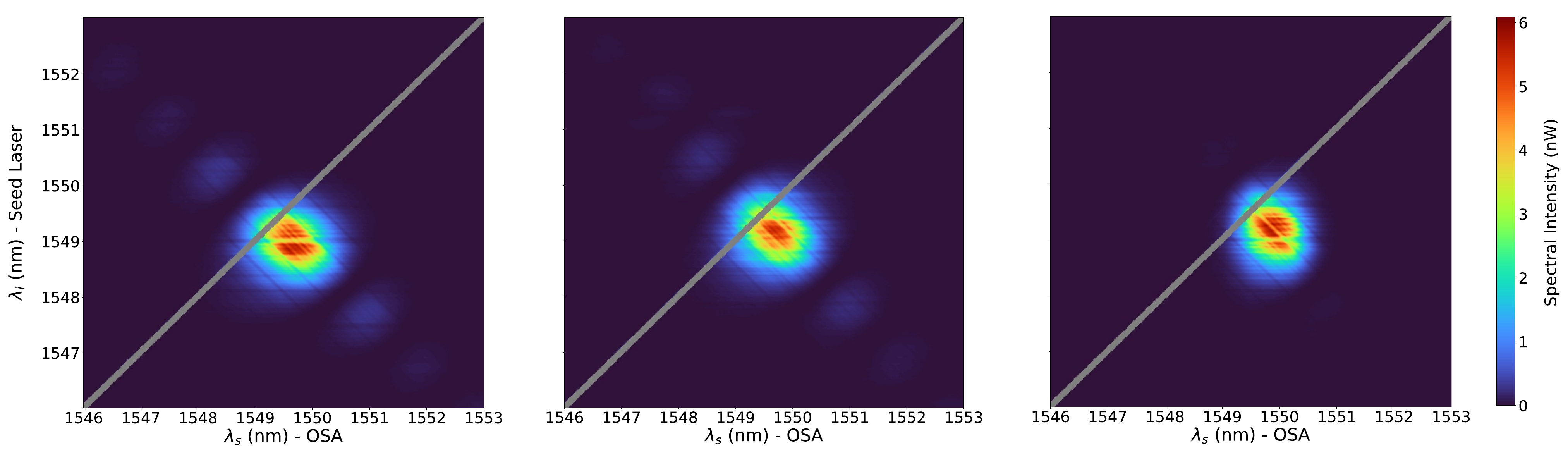}
\caption[JSI_results]{Experimental results for the joint spectral intensity measurement for (left) unfiltered top, (middle) unfiltered bottom and (right) filtered top layers. The heat map represents the power intensity for different combinations of signal and idler wavelengths (x and y axis, respectively). The grey region corresponds to the seed spectrum that was leaked due to the limited extinction ratio of the PBS. The data was collected by sweeping the wavelength of the seed laser, in steps of 20~pm, and analysing the respective idler spectrum, averaged over 20 scans with a resolution of 20~pm.
}
\label{fig:jsi_results}
\end{figure*}

The setup used to measure the JSI is illustrated in Fig.~\ref{fig:jsi} b) and the results are shown in Fig.~ \ref{fig:jsi_results} a) and b) for both the top and bottom layers with no filters. Apart from the central Gaussian mode, we can observe some side bands that reveal the aforementioned correlations between the spectrum of the signal and of the idler. It is also clear that, unsurprisingly, both layers unveil the same spectrum, with a 1.3~nm wide Gaussian mode centered between [1549, 1550] nm. For this reason, we decided to use filters with that same bandwidth (1.3~nm), such that we remove most of the side mode contributions while not compromising the rate too much. The resulting filtered spectrum can be observed in Fig.~\ref{fig:jsi_results} c), depicting a clear gain in spectral purity, compared to the unfiltered case.

\subsection{Hong-Ou-Mandel Measurement}
\label{sec:HongOuMandel}
\begin{figure}[H]
\centering
\includegraphics[width=0.5\columnwidth]{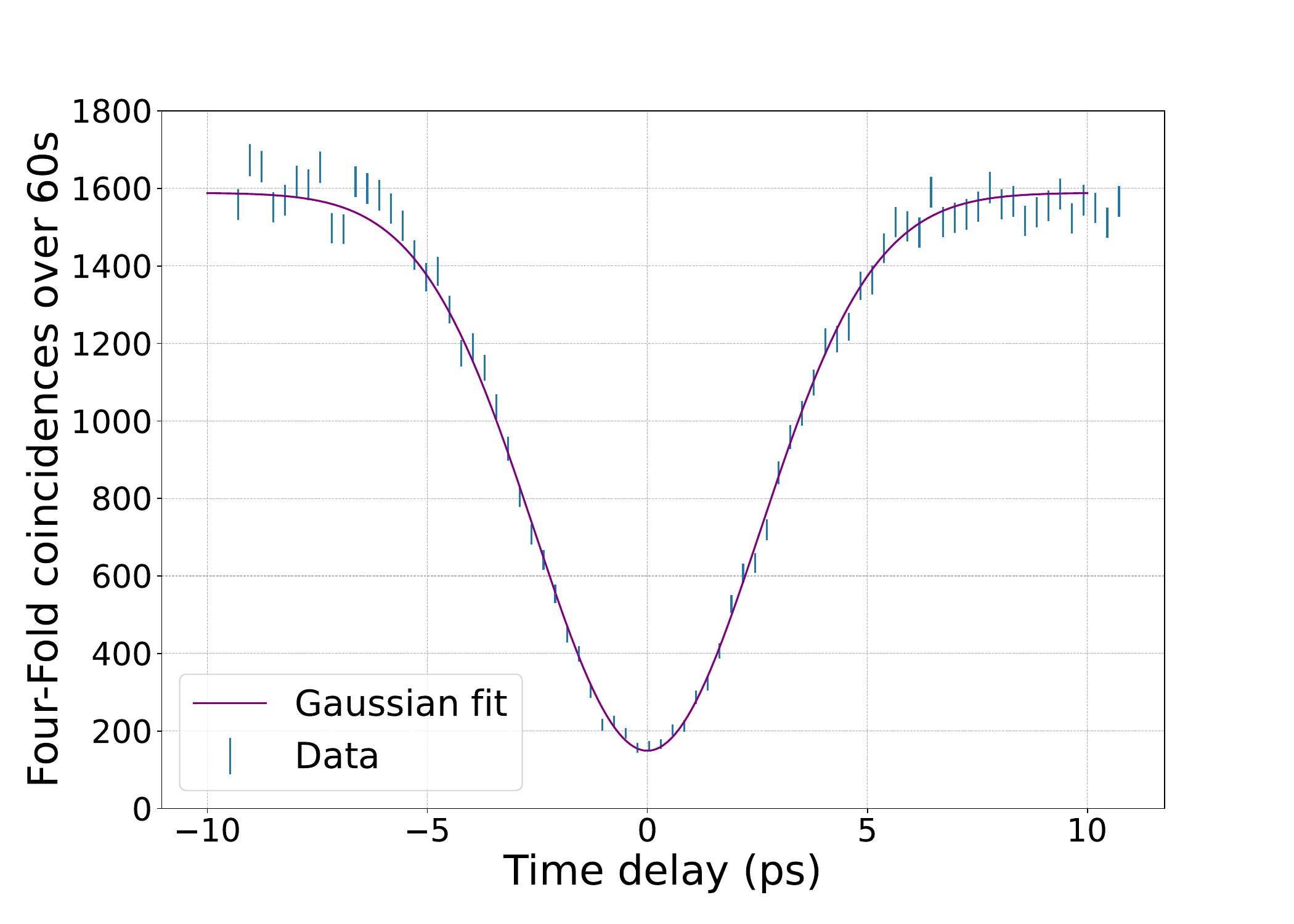}
\caption{Hong-Ou-Mandel dip. The solid purple line is the Gaussian fit to the experimental data (blue points), from which we could infer a visibility of $V=(90.62\pm0.79)\%$ and a coherence time of $\tau = (2\sqrt{2 \textup{ln} 2}) \sigma = (6.03 \pm 0.07)$ ps.}
\label{fig:hom}
\end{figure}

Guaranteeing that the idler photons temporally overlap in the fusion station's FPBS is a key challenge that highly influences the fidelity of our states. A Hong-Ou-Mandel (HOM) measurement can be used to determine how much the coupling of the top idler photon (see Fig.~{\ref{fig:source_setup}}) needs to be delayed at the same time as estimating the visibility of the interference. Figure~\ref{fig:hom} depicts the experimental results, from which we derive a visibility of 90.62$\%$.

\subsection{Unitary Compensation}
\label{sec:app:uc}
As mentioned before, when quantum states are encoded in the polarisation degree of freedom of propagating photons, they suffer from local rotations as they propagate in single-mode fibers due to their birefringence. This means that, in our setup, if we want to have a GHZ, specifically in the form $\ket{GHZ}=(\ket{HHHH}+\ket{VVVV})/\sqrt{2}$, as required by many quantum communication protocols, we need to compensate for these SM fiber rotations.

Considering that any local unitary can be seen as a universal unitary in the SU(2) group, i.e., that it can be written as a function of three independent parameters on the hypersphere $S^3$, it was shown in Ref.~\cite{SIMON1990165} that it is possible to implement any optical local unitary with a set of two QWPs and one HWP, expressed as:

\begin{equation}
\label{eq:unitarycompensation}
U(x_0, x_1, x_2)=\mathrm{QWP}\left(x_0\right) \operatorname{HWP}(x_1) \mathrm{QWP}\left(x_2\right),
\end{equation}
\noindent where QWP$(x)$ (HWP$(x)$) refers to the Jones matrix representation of a quarter-wave plate (half-wave plate) with an angle $x$ between its slow axis and the horizontal direction.

Additionally, if we have a GHZ state up to local unitary transformations, it means that, apart from a relative phase, we require as many unitary transformations as the number of qubits that need to flip, in order to arrive at its desired form. If we also consider that a state resulting from three qubit flips on a GHZ state is always equivalent to some other state with a single qubit flip, we conclude that we require a maximum of two local unitary transformations to arrive at the true GHZ state. However, since there are many combinations of two out of four qubit flips that can occur, depending on the unitary $U_{\text{fiber}}$, we guarantee that we can compensate for all possible rotations by using three sets of QWP-HWP-QWP, as illustrated in Fig.~\ref{fig:source_setup}.

By employing an optimization method we can determine which unitary transformation,  $U^{\dagger}_{\text{fiber}}\cdot\rho\cdot U_{\text{fiber}}$ - of the form $U_{\text{fiber}}=\mathds{1}\otimes U_1 \otimes U_2 \otimes U_3$ - we need to apply, in order to maximize the fidelity of the density matrix, $\rho$ - estimated with QST - with respect to the GHZ state, $\rho_{GHZ}$. We can then decompose $U_{\text{fiber}}$ into sets of angles, according to Eq.~\ref{eq:unitarycompensation}, for the purpose of adjusting the configuration of the waveplates assigned for the unitary compensation.

\end{document}